\documentclass[showpreprint,prb,a4paper,twocolumn,amsmath,amssymb]{revtex4}

\usepackage {times}
\usepackage{graphicx}
\usepackage{dcolumn}
\usepackage{bm}

\begin{document}

\preprint{cond-mat/0406620}

\title{Critical currents in vicinal YBa$_2$Cu$_3$O$_{7-\delta}$ films}

\author{J. H. Durrell}
\email{jhd25@cam.ac.uk}
\author{G. Burnell}
\author{Z. H. Barber}
\author{M. G. Blamire}
\author{J. E. Evetts}
\affiliation{Department of Materials Science and Metallurgy,
University of Cambridge, Pembroke Street, Cambridge, CB2 3QZ, UK.}

\date{July 9, 2004}

\begin{abstract}
Most measurements of critical current densities in
YBa$_2$Cu$_3$O$_{7-\delta}$ thin films to date have been performed
on films where the \textit{c}-axis is grown normal to the film
surface. With such films, the analysis of the dependence of $j_c$
on the magnetic field angle is complex. The effects of extrinsic
contributions to the angular field dependence of $j_c$, such as
the measurement geometry and disposition of pinning centres, are
convoluted with those intrinsically due to the anisotropy of the
material. As a consequence of this, it is difficult to distinguish
between proposed FLL structure models on the basis of angular
critical current density measurements on \textit{c}-axis films.
Films grown on mis-cut (vicinal) substrates have a reduced
measurement symmetry and thus provide a greater insight into the
critical current anisotropy. In this paper previous descriptions
of the magnetic field angle dependence of $j_c$ in
YBa$_2$Cu$_3$O$_{7-\delta}$ are reviewed. Measurements on
YBa$_2$Cu$_3$O$_{7-\delta}$ thin films grown on a range of vicinal
substrates are presented and the results interpreted in terms of
the structure and dimensionality of the FLL in
YBa$_2$Cu$_3$O$_{7-\delta}$. There is strong evidence for a
transition in the structure of the flux line lattice depending on
magnetic field magnitude, orientation and temperature. As a
consequence, a simple scaling law can not, by itself, describe the
observed critical current anisotropy in
YBa$_2$Cu$_3$O$_{7-\delta}$. The experimentally obtained
$j_c(\theta)$ behaviour of YBCO is successfully described in terms
of a kinked vortex structure for fields applied near parallel to
the \textit{a-b} planes.
\end{abstract}

\pacs{72.33Ft}
\maketitle
\section{\label{sec:1}Introduction}
As in all Type-II superconductors it is the structure and pinning
behaviour of the Abrikosov lattice of flux vortices that
determines the critical current properties of
YBa$_2$Cu$_3$O$_{7-\delta}$ (YBCO). A comprehensive understanding
of this is therefore an essential prerequisite to the development
of technologically useful applications of this material.
Low-temperature superconductors are either isotropic or slightly
anisotropic \cite{kog92}. In contrast the more recently discovered
high temperature superconductor (HTS) materials are all layered,
strongly anisotropic materials. Superconductivity is associated
with the cuprate planes which lie in the \textit{a-b} planes of
these materials. For the case of YBCO the Ginzburg-Landau
anisotropy parameter, $\gamma$, is ~5-7, whereas in
Bi$_2$Sr$_2$CaCu$_2$O$_{8+x}$ (BSCCO 2212) the value is $\sim200$.
This large difference has been attributed to the cuprate chains
found along the \textit{b}-axis between the cuprate planes in YBCO
\cite{cava90,hou94,tall95} which may exhibit superconductivity.

For the more strongly anisotropic HTS materials it is reasonable
to start by approximating them as two-dimensional superconductors
with purely Josephson coupling between the superconducting layers
\cite{cle98}. This is the two dimensional superconductor described
by Lawrence and Doniach \cite{law70}. This approach does not
suffice for the case of YBCO. Indeed, over a wide range of applied
field angles it appears that the flux lines in YBCO are the
elliptical vortices which would be expected from anisotropic
Ginzburg-Landau theory \cite{bla92,her97}. For fields applied
nearly parallel to the \textit{a-b} planes Blatter \textit{et al.}
\cite{bla94} have predicted that there is a transition to a kinked
vortex state where the contiguous vortex lines consist of
alternating vortex string and pancakes segments parallel and
perpendicular to the \textit{a-b} planes. Measurements on single
crystals indicate that the vortex lines only fully 'lock-in' to
the planes when the field is within approximately $0.2^\circ$ of
the \textit{a-b} planes \cite{Zhuk99_1}. The fully locked-in state
will only be seen therefore in very perfect crystals.

The variation of the structure of individual flux lines as the
magnetic field direction changes with respect to the \textit{a-b}
planes means that no single scaling law will describe the magnetic
field angle dependence of the superconducting properties of YBCO.
The presence of anisotropic pinning centres will further
complicate the observed behaviour. In order to elucidate the
several contributions to the observed dependence of critical
current on applied magnetic field angle it is necessary to reduce
the measurement symmetry, ideally by arranging that the Lorentz
force is not directed along a crystallographic axis. In this study
we achieve this aim by employing YBCO thin films grown on mis-cut
(vicinal) substrates. Thin films grown on single crystal
substrates are a convenient experimental system especially since
the step-flow growth favoured on vicinal substrates can lead to a
very clean microstructure.

\section{\label{sec:2}Theory}

There has been extensive earlier work on the angular dependence of
critical currents in thin film YBCO. However these studies have
all employed \textit{c}-axis orientated films which greatly
complicates the interpretation of the data obtained.

The first report of the nature of the variation of critical
current with applied magnetic field angle was by Roas and
co-workers \cite{roa90}. Although the angular resolution of their
data was poor they identified the two most prominent features
found in the $j_c(\theta)$ behaviour of YBCO, where $\theta$ is
the angle between the applied field and the \textit{a-b} planes.
When the field is applied parallel to the \textit{a-b} planes they
observed an 'intrinsic pinning' peak which was attributed to
pinning by the layered structure of the superconductor.
Additionally a smaller peak was seen when the field was aligned
parallel to the \textit{c}-axis. This peak was associated by Roas
\textit{et al.} with pinning by twin planes. It is important to
note that in their experiment the field was rotated in a plane
perpendicular to the current so as to keep $j \wedge B$ maximized.
If the field is swept in a plane containing the current direction
the peak in the critical current observed at $\theta=0$ is due to
the 'force-free' \cite{cam72} effect.

It is immediately useful to note that these three peaks arise for
entirely different reasons. The 'intrinsic' pinning peak is simply
due to the anisotropy of the superconductor itself and the
'force-free' peak arises as the Lorentz force $j \wedge B$ tends
to zero. It has since been shown that the primary source of
pinning in \textit{c}-axis films is due to dislocations at the
edge of growth grains \cite{dam99}. It is these dislocations that
lead to the \textit{c}-axis peak since they are most effective as
pinning centres when the applied field is aligned with them. While
the 'intrinsic' peak will not be expected to be sample dependent,
the \textit{c}-axis peak will depend on the film microstructure.
This is supported by the wide variation in the prominence of the
\textit{c}-axis peak seen in different samples.

The nature of the 'intrinsic' peak is not straightforward. Similar
behaviour is observed in BSCCO 2212 which can be quite
satisfactorily described as a two dimensional superconductor. In
BSCCO 2212 the flux lines take the form of stacks of pancake
vortices localized within the \textit{a-b} planes \cite{cle91}. A
consequence of this is that any angular dependent behaviour can be
taken as depending entirely on the \textit{c}-axis component of
the applied field, the 'Kes law' \cite{kes90}. In small fields the
vortex structure does, however, show more complex behaviour
\cite{kosh99}. The 'Kes law' has been found to agree very well
with experiment in BSCCO 2212 \cite{sch91}. Although Jakob
\textit{et al.} \cite{jak93_2} reported that this scaling law also
worked well for the case of YBCO, close examination of their data
shows that while the fit for BSCCO is indeed very good, the
correlation is much less satisfactory for YBCO. Moreover a
consequence of the Kes law is that BSCCO 2212 will not exhibit a
'force-free' effect \cite{cam72} as the local field is not aligned
with the current direction, whatever the orientation of the
external field . This is experimentally observed as constant
critical current when the applied magnetic field is rotated about
the \textit{c}-axis \cite{dur00} with a constant angle between the
field and the \textit{a-b} planes. YBCO does exhibit such an
enhancement of critical current in the 'force-free' geometry
\cite{rav94,her97,dur99}. YBCO cannot, therefore, be treated as
approximating to a two dimensional superconductor.

Tachiki and Takahashi \cite{tac89} considered the effect of
layering in YBCO by noting that for $T < $70 K the core size for a
vortex lying along the \textit{a-b} plane is less than the
inter-planar spacing. They extended their model to cover
orientations other than aligned along the \textit{a-b} plane by
envisaging kinked vortex lines consisting of locked-in vortex
segments connected by short lengths crossing the cuprate planes
\cite{tac89_2}. Although some authors found their data broadly
fitted their predictions \cite{nis91,nis91_2,aom94} none found the
predicted lock-in plateau \cite{fei90}. This is in all likelihood
because the lock-in angle is small $~0.2^\mathit{o}$ and is
suppressed by only slight imperfections in the crystal structure.
The lock-in effect is however seen in high quality single crystals
\cite{Zhuk99_1}.

The intrinsic pinning peak does not disappear at higher
temperatures where the vortex core is larger than the interlayer
spacings. If the superconductor is simply considered as being
anisotropic, Ginzburg-Landau theory predicts that the vortex cores
will become progressively more elliptical as the field is tilted.
This may be described by considering an angle dependent mass
anisotropy parameter \cite{kog92,bla92} $\epsilon_{\theta}$. This
parameter is given by $\epsilon_{\theta}^2 = \epsilon^2 cos^2
\theta + sin^2 \theta$ where $\epsilon$ is the mass anisotropy
parameter, $\epsilon=1/\gamma=\sqrt{\frac{m_{ab}}{m_c}}$. From
this, scaled versions of various experimental parameters may be
obtained, for example $B_{c2}(\theta)=B_{c2}/\epsilon_{\theta}$.
For the case of a tilted vortex line parallel to the \textit{a-b}
planes the vortex core size in the \textit{c} direction is reduced
to $\epsilon_{\theta}\xi_{ab}$, where $\xi_{ab}$ is the G-L
coherence length in the \textit{a-b} planes. If it is assumed that
the smaller core is more effectively pinned, then an 'intrinsic'
pinning peak will still be found. The expected angular dependence
for $j_c$,
\begin{equation}
\label{eq:1} j_c(B,\theta)=j_c(\epsilon_\theta
B,\theta=90^\mathrm{o})
\end{equation}
is not straightforward as, in general, $j_c$ does not depend on
$B$ in a simple fashion. Divergence from this expected behaviour
at a particular angle, $\theta$, can however reasonably be used to
infer the existence of anisotropic pinning, as has recently been
discussed by Civale \textit{et al.} \cite{civ04}

Blatter \textit{et al.} \cite{bla94} reconciled these various
ideas by proposing the existence of a cross-over from a lattice of
conventional rectilinear, but anisotropic, Abrikosov vortices to a
regime of kinked vortices, similar to the Tachiki and Takahashi
model. In the kinked regime pancake vortices in the cuprate planes
are linked by Josephson string vortices in the interlayers. The
pancake vortices have a fully developed normal core of extent
$\xi_{ab}$. The vortices laying between the cuprate planes are
Josephson in nature and do not exhibit full suppression of the
superconducting order parameter.

Blatter introduced two critical angles, $\theta_1$ and $\theta_2$.
Defining $\theta=0$ to be when the applied field is aligned with
the cuprate planes for $\theta>\theta_1$ the flux lines are
conventional straight Abrikosov vortices. For
$\theta_1>\theta>\theta_2$ the flux vortices are distorted towards
the kinked state whilst for $\theta<\theta_2$ the kinked state is
fully formed. They express $\theta_1$ as
\begin{equation}
\label{eq:2} \tan(\theta_1)=\frac{d}{\xi_{ab}(t)}
\end{equation}
$t$ is the reduced temperature ($T/T_c$), $d$ is the interlayer
spacing and $T_c$ is the superconducting transition temperature.
The second critical angle $\theta_2$ is expressed as
\begin{equation}
\label{eq:3} \tan(\theta_2)=\epsilon
\end{equation}
Above a certain critical temperature $T_{cr}$ the transition is
expected to be suppressed and rectilinear Abrikosov vortices are
seen for all orientations of applied magnetic field. This critical
temperature is defined as being that at which
$\xi_{ab}(T_{cr})=d/\sqrt{2}$. In YBCO these values are
\cite{ber97} $\theta_1 \sim 35^\circ$ and $\theta_2 \sim 11^\circ$
for $t$=0. $T_{cr}$ is predicted to be about 80 K. From Eq.
\ref{eq:3} we may deduce that $\theta_2$ will depend only weakly
on temperature since $\epsilon=\xi_c/\xi_{ab}$ and both $\xi_{ab}$
and $\xi_c$ have a similar temperature dependence.

As would be expected, the Josephson string vortices localized
within the \textit{a-b} planes are more weakly pinned than the
pancake vortices \cite{dur99}. In spite of this, a vortex
channelling effect cannot be seen in \textit{c}-axis films since
there is never a component of the Lorentz force directed in the
weakly pinned direction. In the reduced symmetry found in
measurements on vicinal films the expected intrinsic vortex
channelling effect is observed \cite{ber97,durrvic03}. As
predicted from Eq. \ref{eq:2} the width of the vortex channelling
effect has been shown to vary with changing superconducting
anisotropy in calcium doped and de-oxygenated YBCO films
\cite{durran03}.

\section{\label{sec:3}Experimental Technique}
Thin films of YBa$_2$Cu$_3$O$_{7-\delta}$ were prepared by pulsed
laser deposition. The films all had thicknesses of between 100 and
200 nm and were grown on single crystal SrTiO$_3$ substrates
mis-cut by an angle, $\theta_v$, towards the (001) direction.
Vicinal films of YBCO prepared on mis-cut substrates typically
exhibit 'step-flow' growth which leads to a terraced structure
with the terraces perpendicular to the miscut
\cite{hu01,mec98,pop01,wel98,maur03}. It should be noted that the
films had a low density of anti-phase boundaries and stacking
faults, this was confirmed by HREM studies \cite{durrvic03}. These
types of defects disrupt the \textit{a-b} planes and make the
interpretation of observed critical current in terms of the FLL
structure and anisotropy impossible. Such defects, when present,
do however strongly increase the observed critical current
\cite{low95,jos99,jos00}. The different growth modes of YBCO films
on vicinal substrates have been recently discussed by Maurice
\textit{et al.} \cite{maur03}

\begin{figure}
\includegraphics{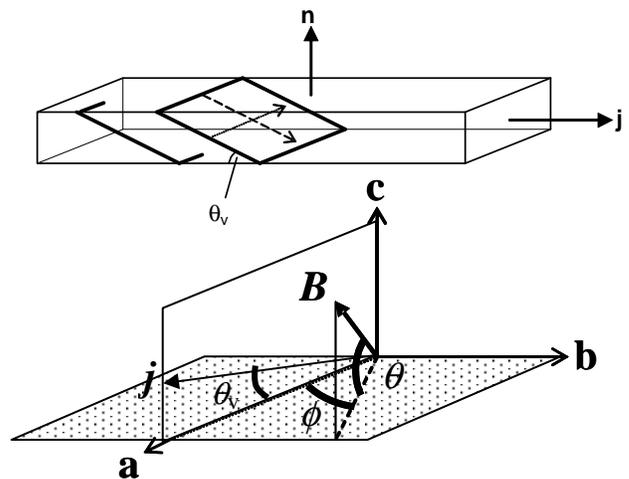}
\caption{\label{fig:2} Geometry of the measurement of a \textbf{T}
track. The upper part of the figure shows the orientation of the
\textit{a-b} planes with respect to the surface of the film. The
mis-cut (vicinal) angle of the substrate is given by $\theta_v$.
The lower part indicates how the tilt angle $\theta$ and rotation
angle $\phi$ specify the orientation of the external magnetic
field. $\theta$=0 is defined as when the field is parallel to the
planes.  At $\theta=\theta_v$ $\phi$=0 the field is aligned with
the current.}
\end{figure}

Current tracks were patterned by photo-lithography and Ar-ion
milling to allow four terminal $IV$ measurements, the tracks were
200x10 $\mu$m between the voltage contacts and were patterned both
perpendicular (\textbf{T}) and parallel (\textbf{L}) to the
vicinal step direction \cite{mec98}. Contacts were prepared by
sputtering Ag/Au bi-layers. The samples were characterized using a
two-axis goniometer \cite{her94} mounted in an 8 T magnet. The
geometry of the measurement, for a \textbf{T} track is shown in
Fig. \ref{fig:2}.

An $IV$ curve was recorded at each set of experimental parameters
($B$,$T$,$\theta$,$\phi$). Critical current values were determined
from these curves using a criterion of 0.5 $\mu$V. The choice of a
voltage criterion to determine critical current, $I_c$ from $IV$
curves is arbitrary, this value was chosen as the lowest practical
value given the noise in the experiment. It was noted that the
overall form of all the observed $j_c$ characteristics did not
change when the saved $IV$ characteristics were reanalyzed with
different voltage criteria.
\section{\label{sec:4}Experimental results}
\begin{figure}
\includegraphics{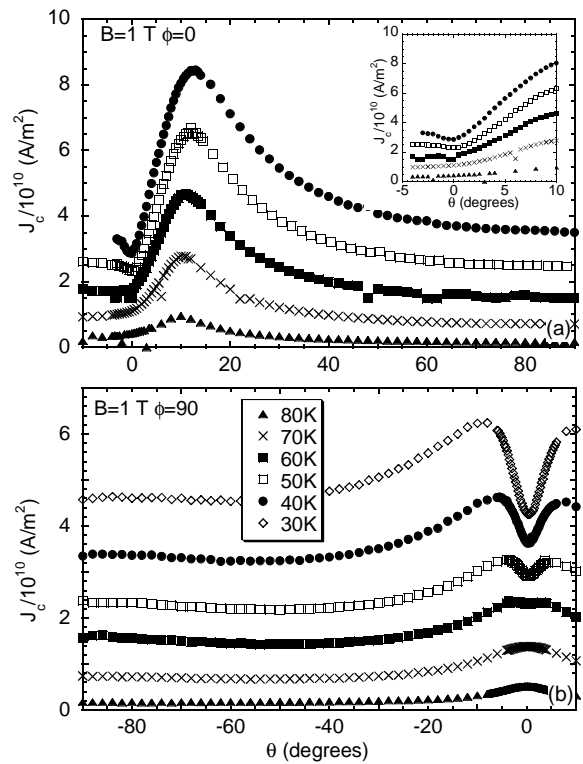}
\caption{\label{fig:3} $j_c$ data taken on a \textbf{T} track
patterned on a YBCO film grown on a 10$^\circ$ mis-cut substrate.
The upper $j_c(\theta)$ plot (a) was obtained with $\phi=0^\circ$
and in a 1 T field, the inset shows an enlargement of the
channelling minimum. The lower plot (b) was obtained with
$\phi=90^\circ$. The vortex channelling effect is visible in both
plots, in (a) the peak arises from the force free geometry whilst
in (b) the peak which is partly suppressed by channelling arises
from intrinsic pinning.}
\end{figure}
In order to provide a fuller characterization of the $j_c(\theta)$
behaviour in vicinal films, and in particular the vortex
channelling effect first reported by Berghuis \textit{et al.}
\cite{ber97}, $j_c(T,B,\theta,\phi)$ studies were performed on
YBCO films grown on substrates with varying mis-cut angles. The
data presented in this article were all obtained on (\textbf{T})
tracks, since it is only in this geometry that the vortex
channelling effect is observed. This occurs since there is no
Lorentz force component in the weakly pinned direction on the
vortex strings in measurements on (\textbf{L}) tracks.

Fig. \ref{fig:3} shows data taken in a 1 T field for a range of
temperatures on a $10^\circ$ vicinal film. The vicinal channelling
minimum is prominent and as expected disappears above 80K
($T_{cr}$) when the flux lines are conventional Abrikosov vortices
for all field orientations \cite{ber97}.

\begin{figure}
\includegraphics{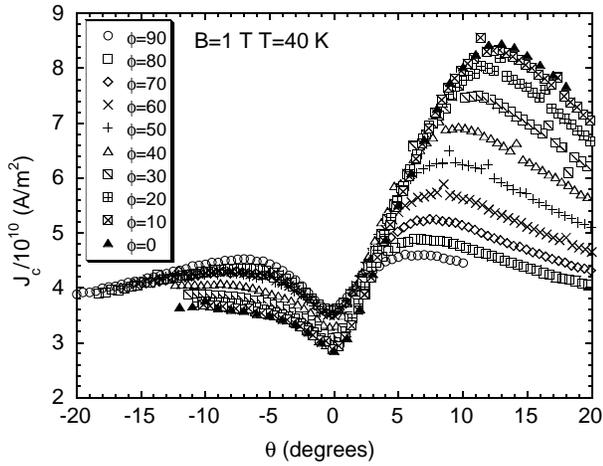}
\caption{\label{fig:4} $j_c(\theta)$ near the vortex channelling
minimum, at 1 T, 40 K and varying values of $\phi$, for a YBCO
film grown on a 10$^{\circ}$ mis-cut substrate.}
\end{figure}
In Fig. \ref{fig:3}a, with $\phi$=0 the channelling minimum,
corresponding to the field being aligned with the \textit{a-b}
planes, is offset by 10$^\circ$ from a peak in $j_c$, which
corresponds to the 'force-free' orientation. This confirms that
the superconducting film indeed has its \textit{c}-axis offset by
10$^\circ$ with respect to the surface normal. In Fig.
\ref{fig:3}b where $\theta$ is swept with $\phi$=$90^\circ$ the
channelling minimum can be seen to be coincident with the
intrinsic peak in $j_c$ which is consequently suppressed. The
channelling minimum is seen for all $\phi$ values, as is shown in
Fig. \ref{fig:4}. The critical current of the minimum can be seen
to be higher at $\phi$=90$^\circ$ than at $\phi$=0$^\circ$ when
$\theta$=0$^\circ$. Although both
$\theta$=0$^\circ$,$\phi$=0$^\circ$ and
$\theta$=0$^\circ$,$\phi$=90$^\circ$ are orientations of the
magnetic field that give rise to vortex channelling, in the former
case the Lorentz force is purely directed along the \textit{a-b}
planes whereas in the latter case there is a component directed
along the \textit{c}-axis as well.

It proved impossible to grow good quality vicinal films with a
clean microstructure for angles greater than $10^\circ$. In such
films no channelling effect was observed. This is probably due to
growth no longer being perfectly epitaxial, this will give rise to
dislocations \cite{low95}, stacking faults and anti-phase
boundaries \cite{jos00} extending through the film. Such defects
will suppress in-plane vortex channelling \cite{durrvic03}.

\begin{figure}
\includegraphics{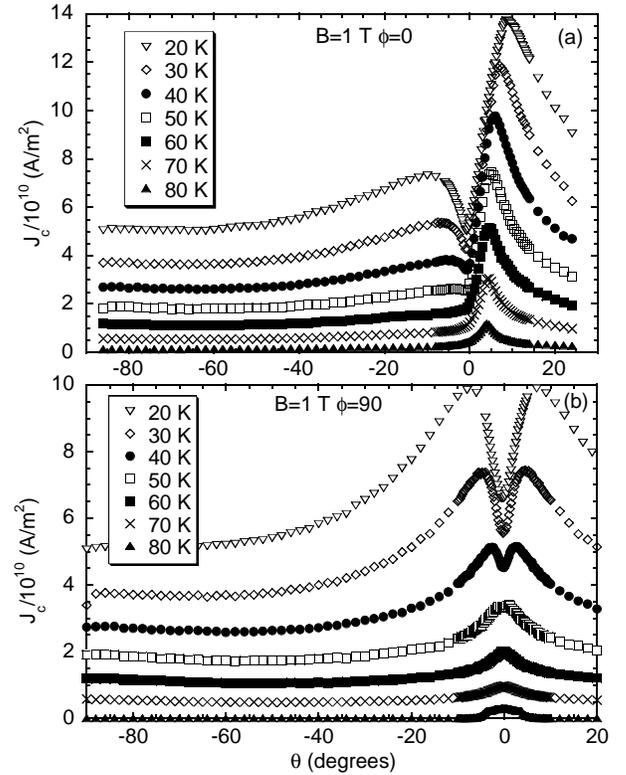}
\caption{\label{fig:5} Critical current, $j_c$ versus field tilt,
$\theta$, data recorded at 1 T and varying temperature on an YBCO
film grown on a 4$^\circ$ mis-cut substrate. In (a) $\phi$ was set
to 0$^\circ$ and in (b) 90$^\circ$.}
\end{figure}

\begin{figure}
\includegraphics{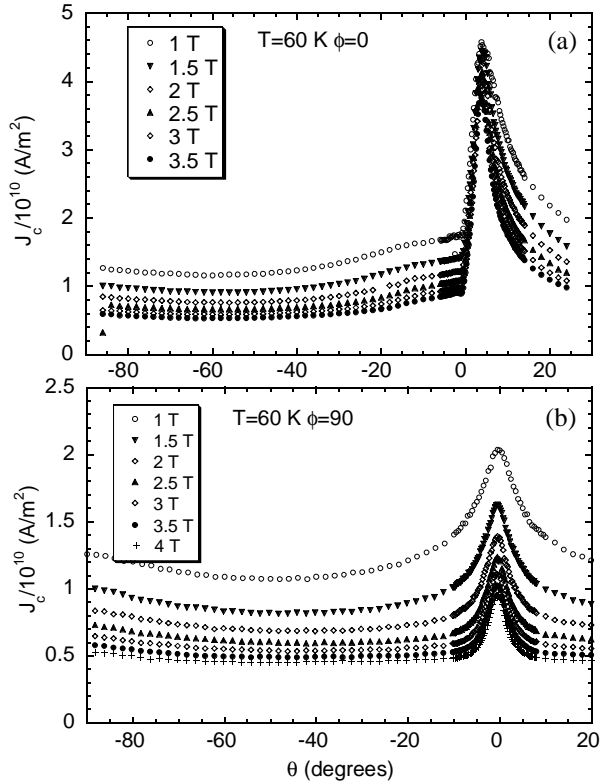}
\caption{\label{fig:6} Critical current, $j_c$ versus field tilt,
$\theta$, data recorded at 60 K and varying field on an YBCO film
grown on a 4$^\circ$ mis-cut substrate. In (a) $\phi$ was set to
0$^\circ$ and in (b) 90$^\circ$.}
\end{figure}

Similar data sets were also obtained on 2$^\circ$ and 4$^\circ$
vicinal films. $j_c(\theta)$ data obtained on 4$^\circ$ films is
shown in Fig. \ref{fig:5} with varying $T$ at 1 T and in Fig.
\ref{fig:6} with varying $B$ at 60 K.
As in the case of the film grown on a 10$^\circ$ mis-cut substrate
we observe  minima due to vortex channelling in both geometries.
As before at higher temperatures the channelling is less
pronounced.

The variation of the $j_c(\theta)$ behaviour with field is shown
in Fig. \ref{fig:6}. In the $\phi$=0$^\circ$ geometry the effect
of vortex channelling is clear and the overall form of the minimum
does not change with varying field, being very sharp. In the
$\phi=90^\circ$ geometry however the channelling effect is not
very pronounced, and the minimum is less distinct. As the
intrinsic pinning peak and the vicinal channelling minimum are
superposed this is a function of the relative strengths of the two
effects.

\begin{figure}
\includegraphics{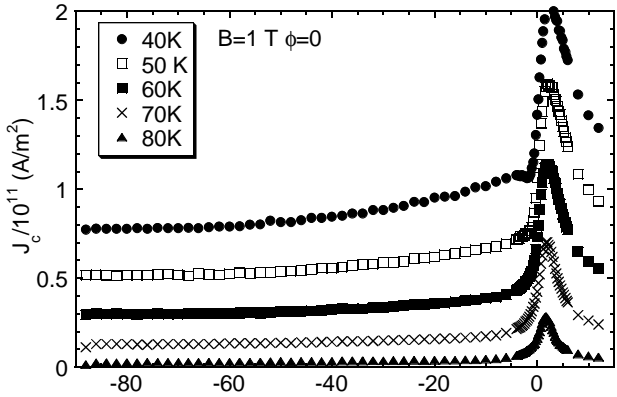}
\caption{\label{fig:8} Critical current, $j_c$ versus field tilt,
$\theta$, data recorded at 1 T and varying temperature on an YBCO
film grown on a 2$^\circ$ mis-cut substrate, $\phi$ was set to
0$^\circ$.}
\end{figure}

Even with a small mis-cut of 2$^\circ$ it is possible to obtain
vicinal growth and consequently observe vortex channelling. This
is shown in Fig. \ref{fig:8}. The close angular proximity of the
vortex channelling minimum and the force-free peak mean that the
critical current increases by a factor of five between
$\theta$=-2$^\circ$ and $\theta$=0. As the minimum due to vortex
channelling and the force free effect are overlapping the point of
maximum $j_c$ is shifted. This is also visible to a lesser extent
in data recorded on the other vicinal angles. Measurements were
not performed on substrates with mis-cuts less than 2$^\circ$, it
is expected however that the vortex channelling effect only
disappears when the mis-cut angle is small enough that the growth
becomes the \textit{c}-axis island growth typically seen in YBCO
thin films.

In summarizing the experimental results it is possible to make
several general observations. The vicinal channelling effect
described in \cite{ber97} in films grown by sputtering has been
reproduced in films grown by pulsed laser deposition. This tends
to suggest strongly that the effect is intrinsic rather than
linked to a particular growth technique. This channelling effect
will therefore be seen in any YBCO sample where the current is not
directed along the \textit{a-b} planes and the crystal structure
is free of defects that disrupt the continuity of the \textit{a-b}
planes \cite{durrvic03}. The channelling effect extends over a
range of miscut angles, the upper limit being due to the
difficulty of getting epitaxial growth on substrates with a large
angle of mis-cut. As previously observed channelling is suppressed
at high temperatures irrespective of the mis-cut angle. Moreover
the effect is most easily seen in the $\phi=0$ configuration where
the channelling minimum is off-set from the 'force-free' peak.
\section{\label{sec:5}Track width}
In this article we seek to identify the extent of the channelling
minimum with the angular range in which there is a cross-over from
rectilinear Abrikosov to kinked string-pancake vortex lines. To
support this argument it is necessary to exclude the possibility
that the width of a particular current track has an effect on the
channelling minimum by changing the length of \textit{a-b} plane
into which a vortex line need align to lock-in. This will be more
pronounced in the $\phi$=90$^\circ$ geometry where the vortex line
must align with the 10 $\mu$m width of the current track.

In the $\phi$=90$^\circ$ geometry the channelling minimum was
measured on a $6^\circ$ vicinal film on a 10 $\mu$m wide track.
The track was subsequently thinned in a focussed ion beam
microscope to a width of 2 $\mu$m. The thinned track was
subsequently remeasured. As gallium ions, used in the milling
process, can have a deleterious effect on YBCO the $T_c$ of the
sample was checked and seen to have remained the same. The milling
was carried out in such a way as to minimise the amount of gallium
spread over the surface of the sample.

\begin{figure}
\includegraphics{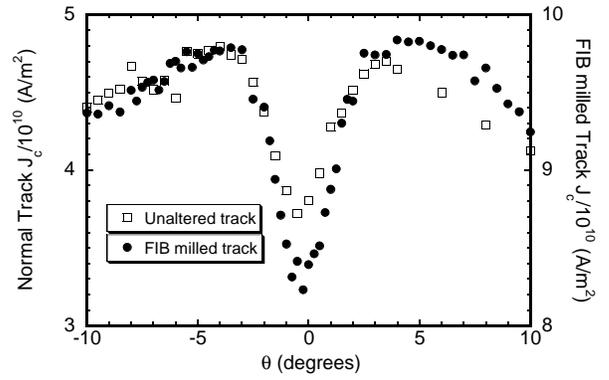}
\caption{\label{fig:FIBres} The vortex channelling minimum
measured on a 10 micron wide track in a 6$^\circ$ vicinal film at
$\phi$=90$^\circ$ before (open squares) and after (closed circles)
thinning to $\approx$ 2$\mu$m using a focussed ion beam
microscope.}
\end{figure}

The results from this investigation are shown in Fig.
\ref{fig:FIBres}. It can be seen that the width of the channelling
minimum is unaffected. The variation in the absolute value of the
critical current may be partly due to uncertainty in the width of
the superconducting channel, the region around a cut made with a
gallium beam will also have suppressed superconductivity due to
gallium implantation.
This measurement clearly demonstrates therefore that the width of
the measurement track does not affect the width of the $j_c$
minimum associated with vortex channelling.

\section{\label{sec:6}Discussion}
The existence of a vortex channelling effect in measurements on
vicinal YBCO films is clear from the series of experiments
presented in this article. In this section the magnetic field
orientation at which cross-over from Abrikosov vortices to the
kinked configuration occurs will be determined. The form of the
channelling minimum will modelled by considering the Lorentz
forces and pinning on the separate vortex elements.

\begin{figure}
\includegraphics{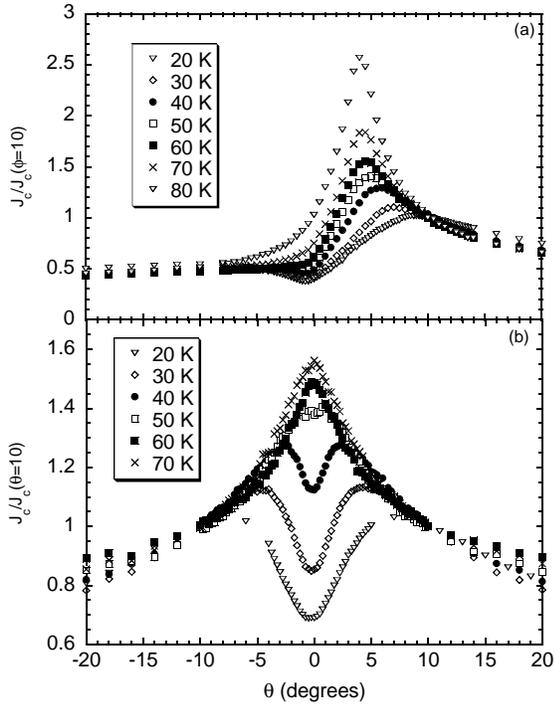}
\caption{\label{fig:4degscale} Data from Fig. \ref{fig:5} plotted
against $j_c/j_c(\theta=10^{\circ})$. ($\theta_v=4^\circ$)}
\end{figure}

As has been discussed previously the scaling law described in Eq.
\ref{eq:1} is not expected to work particulary well over the
entire angular range where the vortices are conventional in
nature. This is primarily due to enhanced pinning at
$\theta=0^{\circ}$ from twins and dislocations. However, away from
both this enhanced extrinsic pinning and the cross-over into the
kinked vortex state the primary contribution to $j_c(\theta)$ is
not unreasonably expected to arise primarily from the anisotropy
of the material. In Fig. \ref{fig:4degscale} the $j_c(\theta)$
characteristic with varying temperature from Fig. \ref{fig:5} is
shown plotted as $j_c(\theta)/j_c(\theta=10^\circ)$. If the
scaling law applies temperature should not change the form of the
rescaled curves.

From the figure this does indeed seem to be the case outside the
range where the behaviour crosses over into the vicinal
channelling regime. In Fig. \ref{fig:4degscale} the region over
which the $j_c(\theta)$ plots diverge has a range of about
$20^\circ$, this is consistent with the prediction of Eq.
\ref{eq:3} since $\theta_2$ depends only on the ratio between the
\textit{a-b} plane and \textit{c}-axis coherence lengths. In Fig.
\ref{fig:4degscale}b all the peaks follow the same form with the
position of the crossover into the channelling minimum appearing
to depend on temperature.

Given that the $j_c(\theta)$ characteristic cannot be described in
terms of the simple G-L scaling equation given in Eq. \ref{eq:1}
it is important to determine if the behaviour near
$\theta=0^{\circ}$ is consistent with kinked vortex lines. If the
vortex lines are indeed kinked it should be possible to predict
the $j_c(\theta)$ behaviour by considering the Lorentz forces and
pinning on the two types vortex segment, vortex strings and
pancakes.

In the $\phi=0^{\circ}$ case the external field direction lies
parallel to the current flow for $\theta=\theta_v$ and is tilted
toward the \textit{c}-axis as $\theta$ is varied. For an
individual rectilinear vortex line the Lorentz force is given by
$j_c \times \Phi_0=f_L$ where $f_L$ is the Lorentz force per unit
length on a vortex line. At the critical current $f_L$ is simply
equal to $f_p$, the pinning force per unit length. This analysis
assumes the pinning forces arise from strong pinning centres. In
this limit the collective pinning volume is that of a single
vortex and it is reasonable to consider pinning forces acting on
individual vortex lines. In the case of a kinked vortex line the
Lorentz forces on and the pinning of the separate segments will be
different. For a kinked vortex line spanning two points
$\textit{l}$ apart and at an angle $\theta$ to the \textit{a-b}
planes there will be a length $l_p=l \sin\theta$ of pancake
elements and $l_s=l \cos\theta$ of string elements. Where the
vicinal angle is $\theta_v$ the Lorentz force on the entire line
per unit length will be (the Lorentz forces in this geometry being
in the same direction on both elements):
\begin{equation}
\label{eq:phi0lorentz} f_L=j \phi_0(\cos\theta \sin\theta_v -
\sin\theta \cos\theta_v)
\end{equation}
Immediately it can be seen that even a kinked vortex line will
experience a force free effect since at $\theta=\theta_v$ $f_L$
will be zero. This is indeed what is experimentally observed,
however experimentally the critical current does not become
infinite. There will therefore be two regimes for angle dependence
of the critical current in the kinked vortex regime. One will
apply where dissipation arises from the vortices depinning, the
second will arise in the regime near the force free orientation
where a different dissipative mechanism must apply.

\begin{figure}
\includegraphics{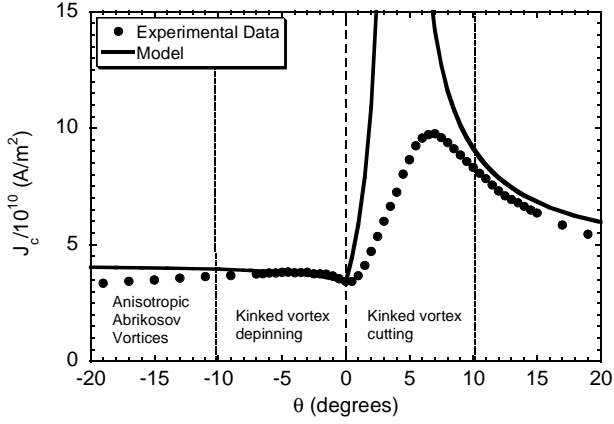}
\caption{\label{fig:model1} Comparison of the model treating the
pinning on individual vortex segments to experimental data in the
$\phi$=0$^\circ$ orientation. The experimental data was taken on a
4$^\circ$ vicinal film at 1 T and 40K. The dotted lines indicate
$\theta_2$ and the dashed line the point at which the direction of
the Lorentz force on the string segments reverses direction.}
\end{figure}

The pinning force available on the vortex lines maybe calculated
by considering pinning on individual string and pancake elements.
The pinning force per unit length of vortex opposing the motion of
the vortices will be given by:
\begin{equation}
\label{eq:phi0pin} f_p=f_{p,str}\cos\theta+f_{p,pc}\sin\theta
\end{equation}
where the subscript 'pc' refers to forces on vortex pancakes and
'str' to those on vortex strings. At $j=j_c$ $f_p=f_L$ so we may
write the following equation for $j_c(\theta)$ combining Eqs.
\ref{eq:phi0lorentz} and \ref{eq:phi0pin}:
\begin{equation}
\label{eq:theta0mod}
j_c=\frac{f_{p,str}\cos\theta+f_{p,pc}\sin\theta}{\phi_0(\cos
\theta \sin\theta_v - \sin\theta \cos\theta_v)}
\end{equation}
The fit of the model to data on a 4$^\circ$ miscut samples at 40K
and 1 T is shown in Fig. \ref{fig:model1}, here $f_{p,pc}$ is
$8.1\textrm{x}10^{-5}$ Nm$^{-1}$ and $f_{p,str}$ is
$4.9\textrm{x}10^{-6}$ Nm$^{-1}$.
The model, with simply two parameters, reproduces the main
features of the experimentally observed behaviour, however it
predicts an infinite critical current when the current is aligned
with the field since, as discussed above, at this point the
Lorentz forces on the string and pancake segments are opposite and
balance out. Three regions may be identified; in the first the
flux lines are rectilinear but anisotropic, in the second the
vortices are kinked but depin conventionally and in the third
regime around the 'force-free' geometry the critical current must
arise from an alternative mechanism.

In this distinct regime, which extends between $\theta=0^{\circ}$
and slightly beyond the 'force free' orientation where
$\theta=\theta_v$, a different mechanism must account for the
finite critical current. In this range the Lorentz forces on the
vortex string and pancakes segments are in opposite directions.
The vortex pancakes are strongly pinned and subject to only a
small force per unit length. It is likely, therefore, that
dissipation occurs through cutting and joining with adjacent
vortices when each string segment is subject to a large enough
force. Flux cutting processes have been described by several
authors, it is important to note that at no time in the phase slip
process does a 'free' vortex end appear
\cite{cam72,blam85,reich04,carr95,grig02}.

If a single vortex string segment is considered it will be subject
to certain Lorentz force, which will be opposed by the elastic
tension in the vortex string as it is deformed and by the pinning
force on the vortex string. When the force on the vortex string
exceeds a certain value, $f_{cut}$ the vortex string cuts and
joins. This force represents therefore the critical current
condition. A similar mechanism has been identified in grain
boundaries in YBCO, which is another system in which vortex
channelling is found \cite{durrgb03}.

Taking a single vortex string in the $\phi=0^\circ$ geometry the
Lorentz force on the vortex string segment will be given by
$f_{L,seg}=j_c \phi_0 d \sin \theta_v \cos \theta  / \sin \theta$.
The length of an individual vortex segment for a particular
magnetic field tilt is given by $d/\sin \theta$, where $d$ is the
spacing between cuprate planes. In YBCO $d$ is one third of the
\textit{c}-axis lattice parameter $\approx$0.39 nm. The available
pinning is given by $f_{p,seg}=f_{cut}+(f_{p,str}d/\sin \theta)$
where $f_{cut}$ is the force required on a single segment for
cutting and cross-joining to occur. Combining these we may write
the following equation for the $\theta$ dependence of the critical
current in the vortex cutting region.

\begin{equation}
\label{eq:fluxcut} j_c=\frac{1}{\phi_0}\left[\frac{f_{cut}
\tan\theta}{d \sin\theta_v}+\frac{f_{p,str}}{\sin \theta_v \cos
\theta}\right]
\end{equation}

\begin{figure}
\includegraphics{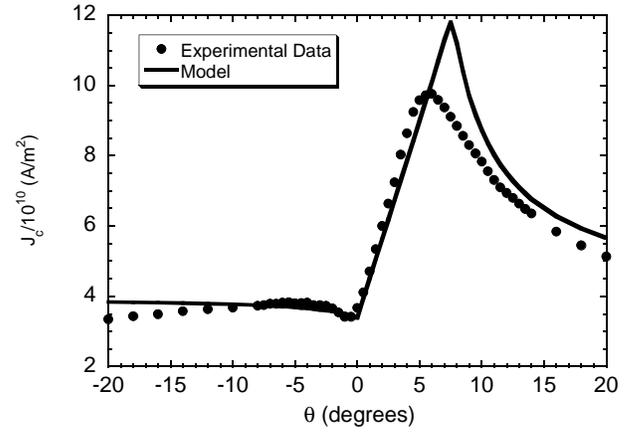}
\caption{\label{fig:fluxcut} The figure shows the behaviour
predicted by extending the model shown in Fig. \ref{fig:model1} to
take account of vortex cutting when the Lorentz forces on the
string and pancake segments are in opposite directions. The
experimental data was taken on a 4$^\circ$ vicinal film at 1 T and
40K.}
\end{figure}

The actually observed critical current behaviour for a kinked
vortex will be the lower of that predicted by Eqs.
\ref{eq:theta0mod} and \ref{eq:fluxcut} for any particular value
of $\theta$. This simply because flux flow occurs if it requires a
smaller Lorentz force than that required for the cutting process.
If the simple assumption is made that the cutting force required
per segment is independent of its length choosing a cutting force
of $3.6\textrm{x}10^{-14}$ N for the 40 K and 1 T data shown in
Fig. \ref{fig:model1} results in a reasonable fit to the
experimental data. This is shown in Fig. \ref{fig:fluxcut}.

It is clear therefore that there is a region between
$\theta=0^{\circ}$ and $\theta\approx\theta_v$ where the flux
flows by cutting and cross joining rather than by depinning entire
vortex lines. Treating the flux cutting force per segment as being
independent of length appears to be a reasonable first
approximation.

\begin{figure}
\includegraphics{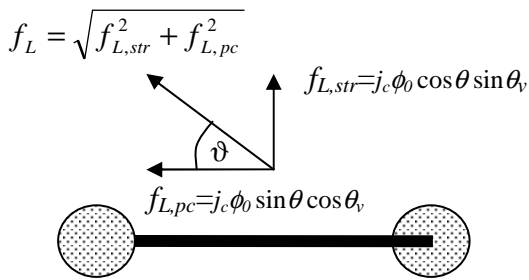}
\caption{\label{fig:phi90dia} Schematic of the forces on a segment
of kinked vortex line, the \textit{c}-axis is perpendicular to the
page. The figure shows, schematically, a vortex string connecting
two vortex pancakes. One pancake will lie in the cuprate plane
above the string the other in the cuprate plane below it. The
angle $\vartheta$ describes the rotation of the resolved force on
the vortex string within the \textit{a-b} planes. Forces directed
outside the \textit{a-b} plane may be neglected as strong
intrinsic pinning restricts the motion of vortex segments in the
\textit{c}-axis direction.}
\end{figure}

A similar approach may be used to model the behaviour of the
$\theta$ dependence of the critical current in the
$\phi=90^{\circ}$ orientation by separately considering the forces
on the string and pancake vortex segments. Here the field is
tilted so that the flux lines are always perpendicular to the
current. The force on the vortex strings will be directed normal
to the plane of the film while the forces on the pancake segments
is directed along the \textit{a-b} planes. The strong intrinsic
pinning means that the component of the force on the vortex
strings directed out out of the \textit{a-b} planes is always
opposed by a an equal pinning force and the forces need only be
resolved within the plane. A section of a tilted vortex line is
shown in Fig. \ref{fig:phi90dia}.
The component of the force on the strings within the \textit{a-b}
plane per unit length is given by $j \Phi_0 \cos\theta
\sin\theta_v$. The force per unit length on the vortex pancakes is
given by $j \Phi_0 \sin\theta \cos\theta_v$, the $\cos\theta_v$
factor arising from the fact the the pancake segments are not
perpendicular to the current. The resulting force per unit length
on the vortex is then:
\begin{equation}
\label{eq:phi90pin} f_L=j_c \phi_0 \sqrt{(\cos\theta
\sin\theta_v)^2+(\sin\theta \cos\theta_v)^2}
\end{equation}.

\begin{figure}
\includegraphics{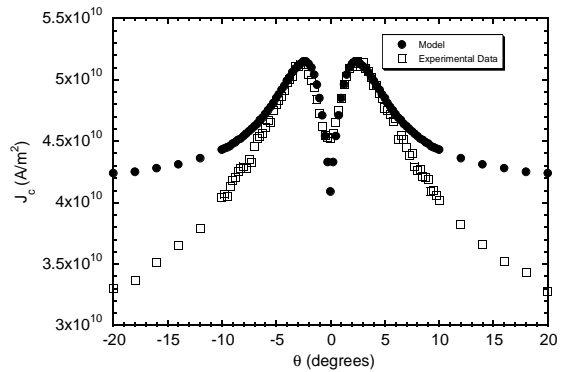}
\caption{\label{fig:model2} Comparison of experimental data with
the model prediction for an external magnetic field in the
$\phi=90^\circ$ geometry.The experimental data was taken on a
4$^\circ$ vicinal film at 1 T and 40K.}
\end{figure}

The pinning force will act in the opposite direction to this force
with the pancake pinning per unit length given by
$f_{p,pc}\sin\theta$. The string pinning force must take into
account that the vortex string is only pinned for movement
perpendicular to its length \cite{dur99} and will be given by
$f_{p,str}\cos\theta \sin\vartheta$. From Fig. \ref{fig:phi90dia}
we can see that orientation of the Lorentz within the \textit{a-b}
plane, $\vartheta$ will be given by
$\tan\vartheta=f_{L,pc}/f_{L,str}$. We may therefore write the
following equation for $j_c(\theta)$:
\begin{widetext}
\begin{equation}
\label{eq:phi90model} j_c=\frac{1}{\phi_0}
\left(\frac{f_{p,pc}\sin\theta}{\sqrt{(\cos\theta
\sin\theta_v)^2+(\sin\theta
\cos\theta_v)^2}}+\frac{f_{p,str}\cos^2\theta
\sin\theta_v}{(\cos\theta \sin\theta_v)^2+(\sin\theta
\cos\theta_v)^2}\right)
\end{equation}
\end{widetext}
Using Fig. \ref{eq:phi90model} it is possible to compare this
model for the $\theta$ dependence of the critical current to
experimental data. A comparison to data obtained on a 4$^\circ$
vicinal film at 40 K and 1 T is shown in Fig. \ref{fig:model2}.
Within the region for which the model is expected to be valid,
$|\theta|<\theta_2$ a good fit to the experimental data is
obtained with $f_{p,pc}=8.1\textrm{x}10^{-5}$ Nm$^{-1}$ and
$f_{p,str}=6.2\textrm{x}10^{-6}$ Nm$^{-1}$. This is the same
pancake pinning force as the $\phi=0^\circ$ geometry but the
required string pinning force is marginally higher. One possible
source of this discrepancy may be that only the force balance in
the \textit{a-b} plane has been considered. The vortex strings are
also subject to a Lorentz force which tends to push them in the
\textit{c}-axis direction, this may increase the pinning force
against movement in the \textit{a-b} plane.
\section{\label{sec:7}Conclusion}
In the high $T_c$ superconductor YBa$_2$Cu$_3$O$_{7-\delta}$ there
is a temperature and magnetic field orientation dependent
crossover from a lattice of rectilinear Abrikosov flux lines to a
lattice of 'kinked' vortices. These kinked vortex lines consist of
vortex pancakes crossing the cuprate planes and Josephson strings
aligned in the charge reservoir layers.

As a consequence of this, 'intrinsic' vortex channelling is
observed providing the superconducting transport current does not
flow parallel to the \textit{a-b} planes. The minima in the
critical current versus magnetic field angle characteristic occurs
not because the vortex lines are 'locked-in' to the \textit{a-b}
planes over a wide angular range but because the transition to the
kinked vortex state changes the angular dependence of the
magnitude of the avaliable pinning force. At the angle where the
vortex lines are expected to align with the planes imperfections
in the films and thermal activation probably ensure that the
vortex lines are never entirely string like. We have shown that
the angular dependence of the critical current in the 'kinked'
range may, to a first approximation, be considered by summing the
Lorentz force on the separate vortex elements and comparing this
to the pinning force available to these elements. The vortex
channelling effect arises because although vortex strings are
strongly pinned for motion parallel to the \textit{c}-axis they
are comparatively weakly pinned for motion in the \textit{a-b}
planes.

In the regime where the forces on vortex pancakes and strings are
opposed it is apparent that a different mechanism serves to limit
the critical current behaviour. We argue that it is cutting and
cross joining of individual vortex string segments that give rise
to flux flow. The vortex pancakes themselves do not flow in this
region.

For magnetic fields well away from parallel to the \textit{a-b}
planes, the properties of the rectilinear vortices may be treated
by using anisotropic Ginzburg-Landau theory to describe how these
elliptical vortices interact with pinning centres. The
$j_c(\theta)$ behaviour in thin films does not, however, scale
using the simple scaling law obtained by Blatter \textit{et al.}
since the strong pinning found in such films is highly
anisotropic.

The complexity of the angular variation of the vortex structure in
YBa$_2$Cu$_3$O$_{7-\delta}$ is due to this material's relativly
low superconducting anisotropy as compared to most other high
T$_c$ superconductors. The consequent cross over transitions in
the vortex lattice make it impossible to apply a single scaling
law to the angular dependence of the critical current.
\begin{acknowledgments}
This work was supported by the Engineering and Physical Sciences
Research council.
\end{acknowledgments}
\bibliography{vicinal}
\end{document}